\documentclass[aip,cha,english,reprint,a4paper]{revtex4-2}

\usepackage{amsmath,amssymb}
\usepackage{graphicx}
\usepackage{xcolor}
\usepackage{hyperref}
\usepackage{physics}
\usepackage{dsfont}
\usepackage{footmisc}
\usepackage{cleveref}
\hypersetup{colorlinks=true,citecolor={blue},linkcolor={blue},urlcolor={blue}}
\usepackage[normalem]{ulem}

\newcommand{\pcsadd}{Center for Theoretical Physics of Complex Systems, Institute for Basic Science (IBS), Daejeon 34126, Republic of Korea}
\newcommand{\ustadd}{Basic Science Program, Korea University of Science and Technology (UST), Daejeon, Korea, 34113}

\begin{document}

\title{Universal Anderson Localization in One-Dimensional Unitary Maps}

\author{Ihor Vakulchyk}
\email{igrvak@gmail.com}
    \affiliation{\pcsadd}
    \affiliation{\ustadd}

\author{Sergej Flach}
\email{sflach@ibs.re.kr}
 \affiliation{\pcsadd}
 \affiliation{\ustadd}
x

\begin{abstract}
   We study Anderson localization in a discrete-time quantum map dynamics 
 in one dimension with nearest-neighbor hopping strength $\theta$ and quasienergies located on the unit circle. We demonstrate that strong disorder in a local phase field 
yields a uniform spectrum gaplessly occupying the entire unit circle. The resulting eigenstates are exponentially localized. Remarkably this Anderson localization is universal as all eigenstates have one and the same localization length $L_{loc}$. We present an exact theory for the calculation of the localization length as a function of the hopping, $1/L_\text{loc}=\left|\ln\left(|\sin(\theta)|\right)\right|$, that is tunable between zero and infinity by variation of the hopping $\theta$.

\end{abstract}

\date{\today}

\maketitle

\begin{quotation}
Discrete-time quantum maps have been established as a powerful testbed for a plethora of quantum and nonlinear phenomena. The map nature of evolution provides remarkable numerical speeds that open the door to previously inaccessible physical regimes, and this has been exploited in research on localization, nonlinear delocalization, soliton formation, Lyapunov spectra properties, and other phenomena. Moreover, maps have eigenvalue spectra which are residing on a compact unit circle space, at variance to continuous-time systems where the relevant eigenvalue space is the infinite real axis. This subtle difference allows to engineer disorder into maps in such a way that the resulting Anderson localization length turns universal within the entire spectrum. Localization length universality is extremely useful when extending the studies by including nonlinearities of many body interactions, since any extended theory builds on the single particle states. Having states which share one localization length instead of an entire spectrum is of great benefit for subsequent analytical treatment. In this work, we propose a simple model that exhibits this feature---Anderson localization universality in a Floquet map---and characterize the phenomenon.
\end{quotation}

\section{Introduction}

Anderson localization (AL) is a phenomenon of prohibited wave transport in various non-interacting quantum systems when placed in an uncorrelated random medium~\cite{anderson1958absence, lee1985disordered, kramer1993localization, lifshits1982introduction}. This is most notably expressed by all the states in the Anderson phase being exponentially localized with the characteristic length $L_\text{loc}$, called the localization length, which is determined by the specifics of the given system.  Since its discovery, AL has been studied in applications to physical systems including photonic crystal waveguides~\cite{sapienza2010cavity}, light~\cite{Schwartz:2007aa,lahini2008anderson}, microwaves~\cite{Dalichaouch:1991aa}, and ultrasound~\cite{WEAVER1990129,Hu:2008aa}. These and many other examples indicate that AL is well understood in Hamiltonian systems, where the spectra of the systems are defined on the real axis and occupy usually a finite strip only, leading to inhomogeneity on the energy axis. This, however, leads to every state being uniquely classified by its energy $E$, and the localization length becomes a function of the state $L_\text{loc}(E)$. A further complication comes from the density of states fading out at the edges of such a spectrum, leading to rare states in some distribution tails, etc.
One notable counterexample is the Aubry--André model~\cite{aubry1980analyticity}, where the underlying potential is pseudo-random (quasi-periodic) and, due to self-similarity, the localization length is universal for all eigenstates~\cite{martinez2018quasiperiodic}. Nevertheless, this does not indicate the universality of the localization length within a continuous portion of the energy spectrum, as its eigenvalue spectrum is fractal.  

Another somewhat less studied situation arises if the dynamics is not continuous in time, thus representing a map. Such systems often result from time-periodic Hamiltonians, i.e. their associated Floquet maps. One notable example is the quantum kicked rotor, in which temporal behavior is strongly connected to AL~\cite{fishman1982chaos} and the corresponding localization length is uniquely defined by the system parameters~\cite{fishman1989scaling}. 
Its downside is the numerical effort to run the map which grows algebraically upon changing the localization length.
Another example is the discrete-time quantum walk (DTQW)~\cite{aharonov1993quantum}, a class of systems discrete in both space and time. Under the application of a random phase field, which uniformly covers the whole complex circle, the localization length has a universal value for any quasi-energy~\cite{vakulchyk2017anderson}. In addition to the emergence of such unique properties, quantum maps present benefits for numerical studies, as the application of a map is much less challenging than numerical integration~\cite{malishava2020floquet, vakulchyk2019wave}.

\begin{figure}
    \includegraphics[width=\columnwidth]{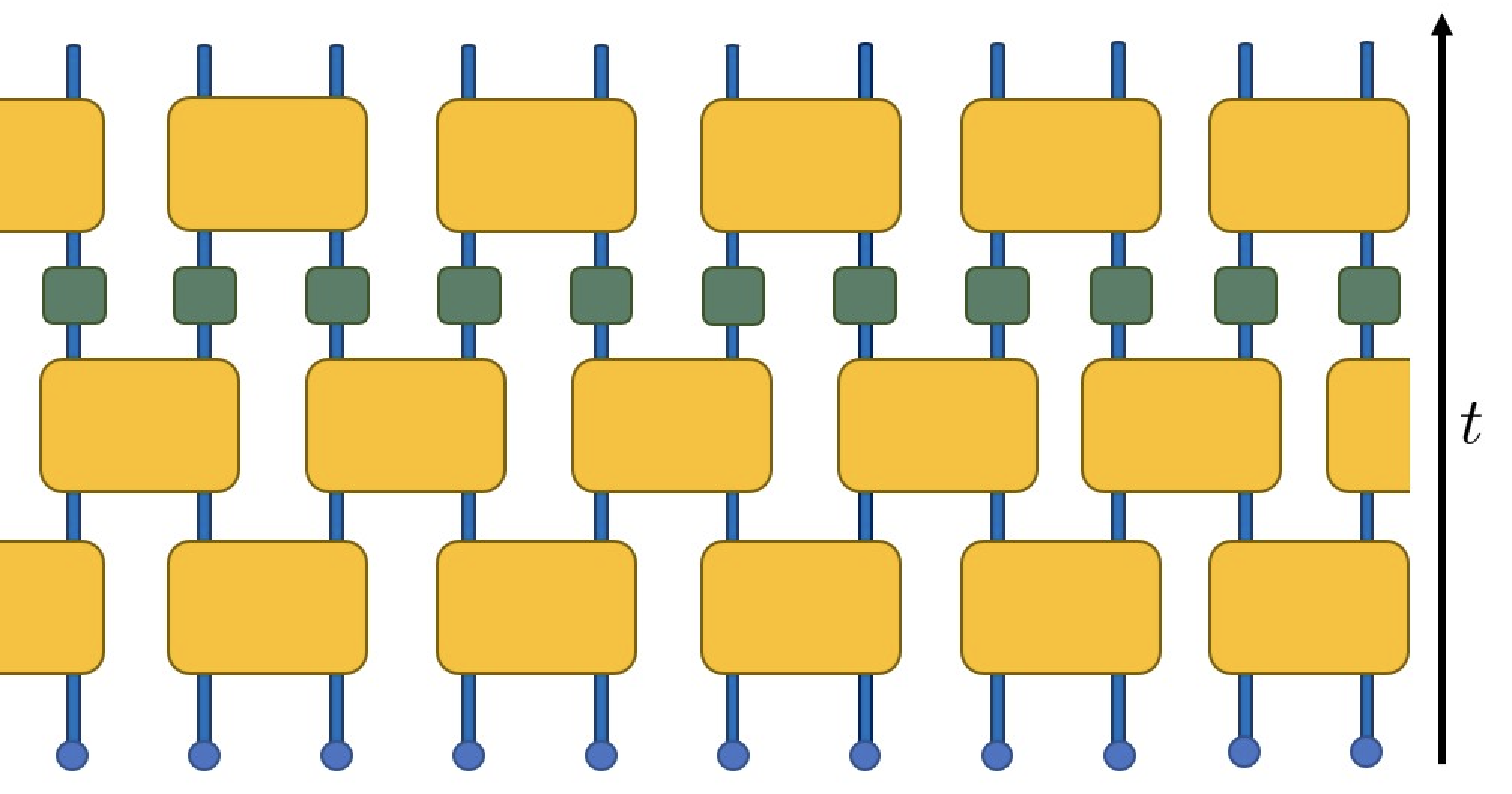}
    \caption{ 
        {Schematic visualization of the quantum map given by~\eqref{eqinit_map}. The vertical arrow represents time, and the boxes indicate the unit cells acted upon. Yellow boxes are two-dimensional unitary matrices acting on a pair of neighbouring sites (\ref{equnitarymatrix}). Green boxes are the spatial disorder acting on each site independently.
}
    }
    \label{figure1}
\end{figure}

In this work, we propose a new model representing a Floquet quantum map in one dimension. Unlike previously studied models, it contains only nearest-neighbor coupling and a gapless spectrum, so the suggested model is in a certain sense a minimal one-dimensional quantum map. Using the transfer matrix approach, we prove that there exist regimes with a universal localization length over the whole spectrum. We provide an exact theory for the calculation of this localization length based on the system parameters. 

\section{Model}

Consider a single particle on a periodic one-dimensional lattice with length $N$ such that each unit cell contains two sites, or in other words, the particle has a spinor degree of freedom. The wave function of such a system reads $\tilde{\Psi}=\{\Psi_{1,+},\Psi_{1,-},\Psi_{2,+},...\Psi_{N,+},\Psi_{N,-}\}$. The Hilbert space can be factorized as a product of the direct space and the spinor space, so $\Psi_{k,\pm}=\ket{k}\otimes\ket{\pm}$. We study a class of time-periodic Hamiltonians that lead to specific Floquet maps defined on the chain, and thus,
\begin{equation}
    \tilde{\Psi}(t+1) = \hat{S}\tilde{\Psi}(t),
\end{equation}
where time $t$ is measured in periods of evolution and $\hat{S}$ is the Floquet map operator. Evolution of such a state $\tilde{\Psi}$ that both a) preserves two sites per unit cell, i.e., is invariant under operation $\sum_{k,\alpha=\pm} \ket{k, \alpha}\bra{k\pm 1, \alpha}$, and b) only involves nearest-neighbor coupling in terms of the Floquet map, $\hat{S}$, can generally (up to the definition of the unit cell) be written in the following form:
\begin{eqnarray}\label{eqinit_map}
    \hat{S} &=& \hat{S}_1 \otimes \hat{S}_2, \nonumber \\
    \hat{S}_1 &=& \sum_{k,\alpha=\pm,\beta=\pm} u_{\alpha\beta}\ket{k,\alpha}\bra{k,\beta}, \nonumber \\
    \hat{S}_2 &=& \sum_{k,\alpha=\pm,\beta=\pm} v_{\alpha\beta}\ket{k,\alpha}\bra{k+l_\beta,\beta}, 
\end{eqnarray}
where $l_\beta = 1$ if $\beta=+$ and 0 otherwise. Thus, two 2x2 matrices $u$ and $v$ contain the parameters defining the system. This evolution may be easily understood in the following way. First, all the spinors are mapped by $u$, then the wave function in the linear form is rotated left, after which all the spinors are mapped by $v$, and finally the wave function is rotated back. See the schematic representation in Fig.~\ref{figure1}.

\begin{figure}
    \includegraphics[width=\columnwidth]{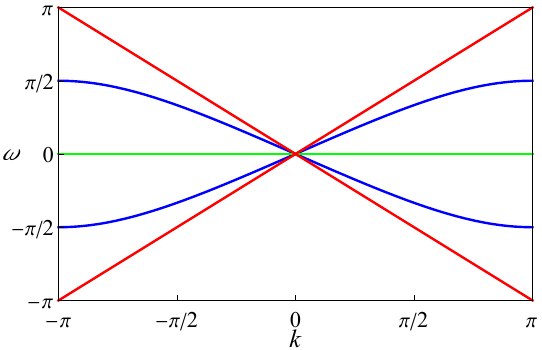}
    \caption{ 
       Band structure from~\eqref{eqdispersion} for $\varphi=\varphi_2=0$ and $\theta = 0,\pi/4,\pi/2$ (red, blue, green).
    }
    \label{figure2}
\end{figure}

One particular well-studied example of such a system is the DTQW. This can be specifically achieved by setting $u$ as a $\mathcal{U}(2)$ matrix that corresponds to a generalized DTQW coin~\cite{chandrashekar2008optimizing}, and setting $v=\sigma^{(x)}$ ($\sigma$ is a Pauli matrix) to realize the so-called shift operation of the DTQW. 

In this paper, we are focused on the case when $u=v$. We take them as the most general unitary matrices,
\begin{equation}\label{equnitarymatrix}
    u=v=e^{i\varphi}\begin{pmatrix}
            e^{i\varphi_1}\cos\theta && e^{i\varphi_2}\sin\theta  \\
            -e^{-i\varphi_2}\sin\theta && e^{-i\varphi_1}\cos\theta
        \end{pmatrix},
\end{equation}
where $\theta,\varphi,\varphi_1,\varphi_2$ are static time-independent parameters characterizing the system, with $\varphi$ being a potential-energy-like local field, and $\theta$ a hopping-like parameter that controls coupling. Given translational invariance, we can find the eigenstates as plain waves given by the ansatz $\psi_k(n) = e^{-i k n} \{\psi_{k,+},\psi_{k,-}\}$. Defining the corresponding eigenvalue of $\hat{S}$ as $e^{i \omega}$, where $\omega$ is the eigenfrequency, we find the dispersion relation
\begin{equation}\label{eqdispersion}
    \cos(\omega-2\varphi) = \cos^2(\theta)+\cos(k-2\varphi_2)\sin^2(\theta).
\end{equation}
Note that $\varphi$ shifts the frequency, $\varphi_2$ shifts the wavenumber, and $\varphi_1$ is irrelevant. The band structure is presented in Fig.~\ref{figure2}. It generally constitutes two bands touching at $2\varphi_2$, where each is exactly $0$ when $\theta = 0$ and straight when $\theta=\pi/2$. 

Without loss of generality we will consider $\phi=\phi_1=\phi_2=0$ leaving $\theta$ as the only relevant control parameter of the ordered map. We will proceed with adding disorder.

\section{Anderson localization}

In this section, we study AL in the case of a random phase field applied to each state component (green boxes in Fig.\ref{figure1}). We modify the translationally invariant evolution defined by~\eqref{equnitarymatrix} by additionally multiplying each component $\Psi_{i,\alpha}$ by $e^{i\phi_{i,\alpha}}$, where $\phi_{i,\alpha}$ are random numbers drawn uniformly from $[-W/2,W/2]$. Therefore the parameter $W$ is the strength of the disorder. This modification is implemented by making the matrix $v$ unit-cell dependent. Note, also, that while we present the result for random phases being independent for each spinor component, the same results apply for unit-cell-only dependent random phases. 

The inverse localization length as a function of the eigenfrequency for different values of disorder strength $W$ is computed numerically using standard transfer matrix methods (see below) and plotted in Fig.~\ref{figure5}. We notice that for the maximum strength of disorder $W=\pi$, the localization length is exactly the same for all the states in the spectrum. This can be explained by the fact that such disorder smears all the frequencies uniformly over the complex unit circle, making them indistinguishable. We will further study this case. 

To calculate the localization length, we use the transfer matrix approach. An eigenstate with an eigenvalue \(e^{i\omega}\) admits the following equation:
\begin{widetext}
\begin{align}
    &\begin{pmatrix}
        \Psi_{n+1,+} \\
        \Psi_{n+1,-}
    \end{pmatrix} = T_n 
    \begin{pmatrix}
        \Psi_{n,+} \\
        \Psi_{n,-}
    \end{pmatrix}, \\
    & T_n =
    e^{2i\varphi}\begin{pmatrix}
        e^{i (\phi_{n+1,+} - \omega)} && e^{-i(\varphi_1-\varphi_2)}e^{i(\phi_{n+1,+} - \phi_{n,-})} \cot{\theta} (1 - e^{i(\phi_{n,-} - \omega)}) \\
        e^{i(\varphi_1-\varphi_2)}\cot{\theta} (1 - e^{i(\phi_{n+1,+} - \omega)}) &&
        \sin^{-2}\theta e^{-i(\phi_{n,-} - \omega)} + \cot^2\theta \left(e^{i (\phi_{n+1,+} - \omega)} - e^{i(\phi_{n+1,+} - \phi_{n,-})} - 1\right)
    \end{pmatrix}. \notag
\end{align}
\end{widetext}

\begin{figure}
    \includegraphics[width=\columnwidth]{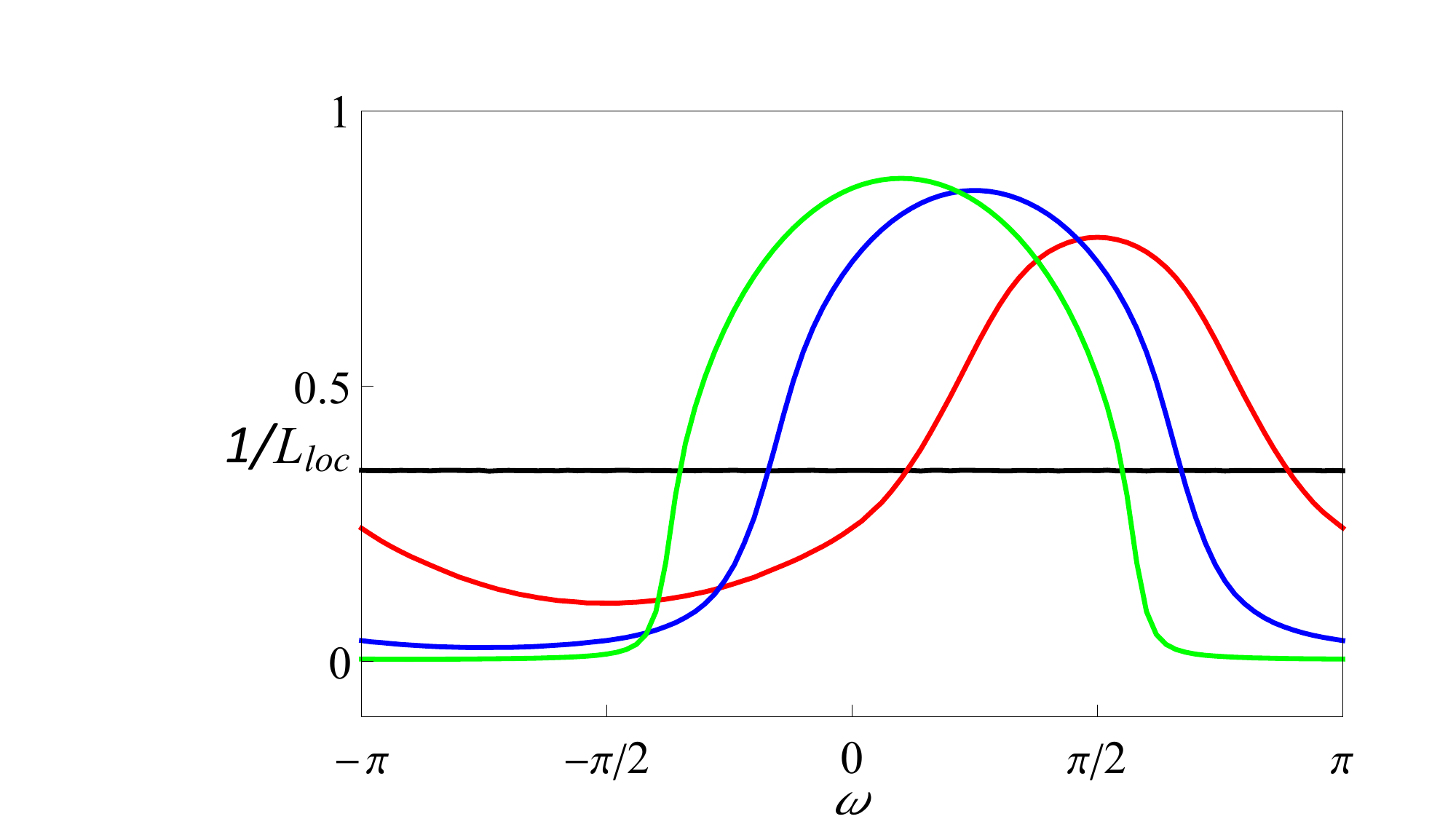}
    \caption{ 
        Inverse localization length versus eigenfrequency of a state for disorder strengths of $W/2=\pi/10,\pi/4,\pi/2,\pi$ (green, blue, red, black). We remind that $\varphi=\varphi_1=\varphi_2=0$.
    }
    \label{figure5}
\end{figure}

Note that the transfer matrix \(T_n\) is always unitary but symplectic if and only if \(\phi_{n+1,+}=\phi_{n,-}\). Also note that the transfer matrix is invariant under \(\phi \rightarrow \phi -\omega\). It immediately follows that for \(W=2\pi\), the ensemble of transfer matrices is completely independent of \(\omega\) as \(\mathcal{U}(S^1)\) is invariant under rotation. Also, note that as the parameter $\varphi$ only enters as a prefactor with an absolute value of 1, it does not affect the localization length. Thus we omit it.
We now introduce a transformation, 
\begin{gather}
    \begin{pmatrix}
        \Psi_{n+1,+} \\
        \Psi_{n+1,-}
    \end{pmatrix} = A_n
     \begin{pmatrix}
        y_n \\
        1
    \end{pmatrix}.
\end{gather}
In order to calculate the localization length, we need to study the behavior of the Riccati variable \(y_n\) at \(n \rightarrow \infty\)~\cite{crisanti2012products}. We can exclude \(A_n\) and write a closed iterative expression for the Riccati variable (\(\omega=0\) as it can be chosen arbitrarily) as follows:
\begin{widetext}
\begin{gather}
    y_{n+1} =  e^{-i(\varphi_1-\varphi_2)}\frac{y_n  e^{i(\varphi_1-\varphi_2)} e^{i\phi_{n+1,+}} + \cot\theta e^{i(\phi_{n+1,+}-\phi_{n,-})} (1-e^{i\phi_{n,-}})}{y_n  e^{i(\varphi_1-\varphi_2)}\cot\theta (1-e^{i\phi_{n+1,+}}) + \sin^{-2}\theta\left(e^{-i\phi_{n,-}}+\cos^2\theta e^{i\phi_{n+1,+}} -\cos^2\theta (1+e^{i(\phi_{n+1,+}-\phi_{n,-})})\right)}.
\end{gather}
\end{widetext}
Rescaling the phase with a constant factor $y_n \rightarrow y_n e^{-i(\varphi_1-\varphi_2)}$ does not change the localization length, so we get rid of these factors, thereby demonstrating that the localization length depends only on the parameter $\theta$ of the model. 
Let us denote $y_n \equiv z$ and and $y_{n+1} = f(z)$.
We then arive at the resulting map \(z \rightarrow f(z)\). To proceed, we transform the space by a shift and rescaling,
\begin{gather}
    z = \cot\theta + z_1/\sin\theta.
\end{gather}
The map \(z_1 \rightarrow f_1(z_1)\) then reads
\begin{widetext}
\begin{gather}
    \label{z1transform}
    f_1(z_1) = \frac{
    z_1 e^{i\phi_{n,-}} (1 - 2 e^{i\phi_{n+1,+}} + \cos (2 \theta))-2 (e^{i\phi_{n+1,+}}-1) \cos\theta}
    {
    2 z_1 e^{i\phi_{n,-}} (e^{i\phi_{n+1,+}}-1) \cos\theta + e^{i\phi_{n+1,+}} \cos (2 \theta)-2 + e^{i\phi_{n+1,+}}}.
\end{gather}
\end{widetext}
This map has a single invariant submanifold \(|z_1|=1\), and thus the stationary distribution of \(z_1\) is located on this unit circle and it is sufficient to calculate the stationary distribution of \(\arg(z_1)\). However, the projection of \eqref{z1transform} on \(\arg(z_1)\) is highly non-trivial and virtually impossible to work with; it is much more common to work with distributions of unbound variables. To make this possible, we transform the \(z_1\) plane using a conformal map, 
\begin{gather}
    z_1 = \frac{z_2 - i}{z_2 + i},
\end{gather}
thus mapping the circle \(|z_1|=1\) onto the real axis. The map \(z_2 \rightarrow f_2(z_2)\) reads
\begin{widetext}
\begin{gather}
    f_2(z_2) = \frac{\tan ^2\left(\frac{\theta}{2}\right) \left(\cos \theta \left(e^{i \phi_1} (z_2-i)+e^{i \phi_2} (z_2+i)\right)+(z_2-i) e^{i (\phi_1+\phi_2)}+z_2+i\right)}{-i \cos (\theta) \left(e^{i \phi_1} (z_2-i)-e^{i
   \phi_2} (z_2+i)\right)+(1+i z_2) e^{i (\phi_1+\phi_2)}-i z_2+1}.
\end{gather}
\end{widetext}
After additionally taking into account that the stationary distribution is located on the real axis, we may get the map for the real part \(x=\Re(z_2)\),
\begin{widetext}
\begin{gather}
    \label{real_map}
    f(x) = \tan ^2\left(\frac{\theta}{2}\right)
    \frac{x \left( \cos(\frac{\phi_1-\phi_2}{2}) \cos\theta + \cos(\frac{\phi_1+\phi_2}{2})\right) +  \sin(\frac{\phi_1+\phi_2}{2}) + \sin(\frac{\phi_1-\phi_2}{2}) \cos\theta}
    {x \left( \sin(\frac{\phi_1-\phi_2}{2}) \cos\theta - \sin(\frac{\phi_1+\phi_2}{2})\right) + 
    \cos(\frac{\phi_1+\phi_2}{2}) - \cos(\frac{\phi_1-\phi_2}{2}) \cos\theta}.
\end{gather}
\end{widetext}
The stationary distribution of \(x\) can now be found as the solution to the integral equation~\cite{derrida1983singular},
\begin{gather}
    \label{int_eq}
    p(x') = \int_0^{2\pi} \frac{\text{d}\phi_1}{2\pi}\int_0^{2\pi} \frac{\text{d}\phi_2}{2\pi}\int_{-\infty}^{\infty} \text{d}x \; p(x)\delta(x'-f(x)).
\end{gather}

\begin{figure}
    \includegraphics[width=\columnwidth]{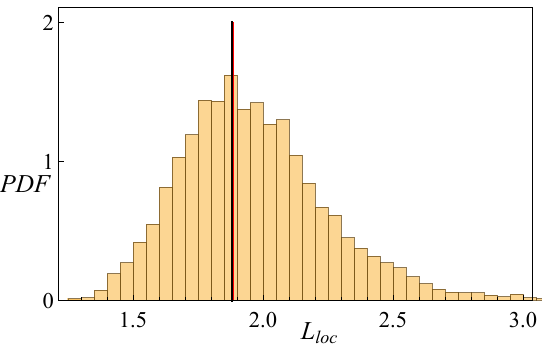}
    \caption{ 
        Localization length for \(W=2\pi, \theta=\pi/5\). The orange histogram shows the probability distribution of localization length for 50 samples with length \(N=2*500\). The red area is the range of values calculated for various eigenfrequencies within \([0,2\pi)\) using numerical transfer matrices with \(10^6\) iterations each. The black line indicates the analytical value from~\eqref{loc_len}. The first moment of the probability distribution is within $0.025$ relative error from the analytic value.
    }
    \label{figure3}
\end{figure}

Unfortunately, the structure of \eqref{real_map} is too complicated to allow a direct solution of \eqref{int_eq}. To get a hint, we notice that the ratios of two random variables are themselves Cauchy distributed under a wide range of circumstances~\cite{pillai2016ratios}. This is in particular always the case for ratios of elliptically symmetric distributions~\cite{arnold1992distributions}. Thus, we may hypothesize that the stationary distribution \(p(x)\) is indeed a Cauchy distribution. To prove this, we note that for any values of random phases \(\phi_1\) and \(\phi_2\), the map from \eqref{real_map} is an element of the real Möbius group \(\text{SL}(2,\mathbf{R})\), and then note that the Cauchy distribution family is closed under this group \cite{mccullagh1996mobius}. Consequently, the stationary distribution of \eqref{real_map} is necessarily a Cauchy distribution. We choose it as follows,
\begin{gather}
    \label{ansatz}
    p(x)=\frac{1}{\pi\gamma (1+x^2/\gamma^2)}, \qquad \gamma = \tan\left(\theta/2\right)^2,
\end{gather}
and we prove that this is the correct choice of the parameter $\gamma$ by showing that it solves~\eqref{int_eq}. Integrating~\eqref{int_eq} over \(x\), we get
\begin{gather}
    \label{int_eq2}
    p(x') = \int_0^{2\pi} \frac{\text{d}\phi_1}{2\pi}\int_0^{2\pi} \frac{\text{d}\phi_2}{2\pi}p\left(f^{-1}(x')\right)/|f'(f^{-1}(x')|,
\end{gather}
where \(f^{-1}\) is the inverse function of \(f\). It generally has more than one value, but not in this case. As we have already proven that the solution is a Cauchy distribution, we only need to confirm the parameter choice, for which it is adequate to integrate at \(x'=0\) only. Substituting~\eqref{ansatz} into~\eqref{int_eq2}, we get
\begin{widetext}
\begin{align}
    & \frac{\sec ^4\left(\frac{\theta}{2}\right)}{4} = \int_0^{2\pi} \frac{\text{d}\phi_1}{2\pi}\int_0^{2\pi}\frac{\text{d}\phi_2}{2\pi}
    \times \notag \\
    & \times \frac{1}{\left(\cos (\theta) \cos \left(\frac{\phi_1-\phi_2}{2}\right)+\cos \left(\frac{\phi_1+\phi_2}{2}\right)\right)^2+\cot ^4\left(\frac{\theta}{2}\right) \left(\cos (\theta) \sin \left(\frac{\phi_1-\phi_2}{2}\right)+\sin
   \left(\frac{\phi_1+\phi_2}{2}\right)\right)^2},
\end{align}
\end{widetext}
which is indeed a true identity. 

As a result of all the above findings, we are finally able to calculate the localization length through

\begin{widetext}
\begin{eqnarray}\label{loc_len}
     \frac{1}{L_\text{loc}} &=& \int \text{d}z \; p(z)\ln\left(|z|\right)  \nonumber \\
     &=& \int_{-\infty}^\infty \text{d}x \; p(x) \frac{1}{2} \ln \left[ \frac{1}{2}\csc (\theta) \left(\frac{4 \left(x^2-1\right) \cot (\theta)}{x^2+1}+(\cos (2 \theta)+3) \csc (\theta)\right)\right] = \left|\ln\left(|\sin(\theta)|\right)\right|. 
\end{eqnarray}
\end{widetext}

The formula
\begin{equation}
 \frac{1}{L_\text{loc}} = \left|\ln\left(|\sin(\theta)|\right)\right|
\label{result}
\end{equation}
is the central result of our work. We show analytically that strong disorder $W=2\pi$ induces Anderson localization with a universal localization length which depends only on the single control parameter $\theta$. As expected, for $\theta \rightarrow 0$ the localization length vanishes as the unitary map decouples all sites. Likewise, for $\theta \rightarrow \pi/2$ the localization length diverges, as even and odd sites on the unitary map chain decouple their evolution.

For usual finite size diagonalization studies, the localization length can be obtained from a subsequent fitting of the tails of the eigenstates. Finite size corrections always apply, leading to a distribution of the measured localization length for $N=500$ as shown in Fig.~\ref{figure3} for $\theta=\pi/5$ 
(note that this distribution has a finite variance, is not Cauchy, and not directly related to (\ref{ansatz})).    Remarkably the analytical result (\ref{result}) yields the value 1.882 which is marked by the black vertical line in Fig.~\ref{figure3} and agrees within 0.025 with the first moment of the measured distribution. Needless to say that the measured distribution will shrink in width upon increasing the size of the used system. 

\begin{figure}
    \includegraphics[width=\columnwidth]{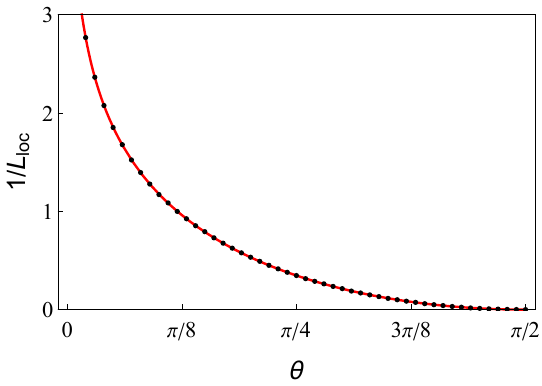}
    \caption{ 
        Inverse localization length versus the hopping-like parameter $\theta$. The red line is the analytical calculation from~\eqref{loc_len}, and the black dots are the numerical results from the transfer matrix approach with $10^6$ iterations for each value of $\theta$.}
    \label{figure4}
\end{figure}

Numerical evaluation of the transfer matrix equations can be easily performed over $10^6$ steps which leads to an excellent agreement with the analytical prediction as shown in Fig.~\ref{figure4}. 

\section{Discussion}

We proposed a minimal one-dimensional Floquet map with only nearest-neighbor coupling of the unit cells and a gapless spectrum. We demonstrated that the Anderson localization length is universal for all the eigenstates and for every value of the quasi-energy on the complex unit circle (in the thermodynamic limit), and depends only on the hopping-like parameter of the model. Furthermore, the localization length can be tuned from $0$ to $\infty$. 

These facts can be widely used in numerical studies. For example, in a recent paper, the authors employed a similar system in the absence of disorder to study Lyapunov spectrum scaling~\cite{malishava2022lyapunov} and conjectured its behavior in the presence of disorder. The universality and ability to control the disordered phase that we showed in this work may benefit subsequent related studies. Further applications may include research into thermalization, ergodization, localization phenomena in the presence of interactions, the interplay between nonlinearity-induced chaos and localization, delocalization due to dissipation, time crystals, and many others. 

Our work serves primarily computational approaches to study disordered systems with the perspective of adding may body interactions \cite{nahum2018operator}. We propose an aesthetically appealing model which produces Anderson localized states with a constant density of states and a single localization length which can be varied from value zero to value infinity by one control parameter. On the other side, the model might be of interest for experimental ramifications, which certainly will bring limitations with time, like the one shown for finite size corrections in Fig.~\ref{figure3}. Likewise our results might be of interest for transport through nanostructures \cite{Celardo:2010vh}, open disordered systems \cite{Sorathia:2012tf}, and qubit management \cite{Tayebi:2016ur}.

\textit{Acknowledgment}.This work was supported by the Institute for Basic Science (Project numbers: IBS-R024-D1, IBS-R024-Y3.). The authors would like to thank Alexei Andreanov and Merab Malishava for their valuable discussions and help, and Joel Rasmussen from Recon for proofreading.

\bibliographystyle{unsrt}
\bibliography{bibl}
\end{document}